# Compression of High Dynamic Range Video Using the HEVC and H.264/AVC Standards

(Invited Paper)

Amin Banitalebi-Dehkordi[1,2], Maryam Azimi[1,2], Mahsa T. Pourazad[2,3], and Panos Nasiopoulos[1,2]

[1]Department of Electrical & Computer Engineering, University of British Columbia, Canada
[2]Institute for Computing, Information and Cognitive Systems, University of British Columbia, Canada
[3]TELUS Communications Inc., Canada
{dehkordi, maryama}@ece.ubc.ca, {pourazad, panos}@icics.ubc.ca

*Abstract*—The existing video coding standards such as H.264/AVC and High Efficiency Video Coding (HEVC) have been designed based on the statistical properties of Low Dynamic Range (LDR) videos and are not accustomed to the characteristics of High Dynamic Range (HDR) content. In this study, we investigate the performance of the latest LDR video compression standard, HEVC, as well as the recent widely commercially used video compression standard, H.264/AVC, on HDR content. Subjective evaluations of results on an HDR display show that viewers clearly prefer the videos coded via an HEVC-based encoder to the ones encoded using an H.264/AVC encoder. In particular, HEVC outperforms H.264/AVC by an average of 10.18% in terms of mean opinion score and 25.08% in terms of bit rate savings.

## I. INTRODUCTION

Recently, media and entertainment industries have been pushing towards High Dynamic Range (HDR) content generation and display systems. The motivation behind these efforts is that HDR delivers dynamic range that is close to what is perceived by the human visual system (HVS) in real life. HVS is capable of perceiving the light approximately at contrast ratio of $10^5$:1 simultaneously in one scene [1]. This range is far beyond the dynamic range that the majority of existing capturing and display devices are capable of providing. Presently, the vast majority of existing consumer cameras and display devices are able to support Low Dynamic Range (LDR) video content with contrast ratio of approximately 100:1 to 1000:1.

For capturing HDR content that embraces the full visible color gamut and dynamic, one solution is to record the scene with different exposure settings simultaneously and then combine the captured LDR videos to create a single HDR scene [2]. Using this method, the information of both under exposed (dark) and over exposed (bright) areas of the scene is preserved. To display HDR content, one approach is to employ a display system which consists of an LDR LCD panel in front and a LDR projector at the back [3]. Such a system is capable of displaying HDR content with contrast ratio of up to 50000:1 [3].

Similar to the LDR content delivery, video compression plays an important role in the HDR broadcasting chain. In the case of HDR, the efficiency of the compression scheme is even more pronounced as HDR content contains significantly more information that the corresponding LDR stream. Each color component in LDR videos is represented by 8 bits (each pixel has three color components and is represented by 24 bits). Instead, each color component in an HDR video stream is represented in a floating-point notation, which is saved in 10 to 16 bits (depending on the file format) [4-6]. Because of that, storing and transmitting HDR video requires much more space and bandwidth than that of LDR. Thus, efficient compression schemes are required for HDR content delivery and storage. To this end, Mantiuk et al. [7] proposed coding HDR content using a scalable approach, where the LDR version of the content is coded as the base layer and the extra information required for converting the LDR content to HDR format is coded separately as the enhancement layer. Both layers are coded based on the existing LDR video coding standards and then, at the decoder side, these layers are combined to generate an HDR video stream [8]. Although the compression performance of this method is not high, backward compatibility with the current LDR displays is ensured. The researchers in [9] modified the structure of the JPEG 2000 standard to achieve compression of HDR images. Mantiuk et al. [10] modified the MPEG 4 video compression standard for HDR video compression applications. They introduced a new luminance quantization scheme, which is optimized HDR content. The above schemes, although proposed for HDR video compression, they utilize existing standards whose rate and distortion performance are optimized for LDR content. To this date there is no video compression standard specifically designed to take advantage of the characteristics of HDR content. The latest LDR video compression standard, HEVC [11], as well as the widely commercially used video compression standard, H.264/AVC [12], support high bit-depth content, but they are designed for coding LDR content and do not take into account the characteristics of the HDR content.

In the absence of a designated compression standard for HDR content, it is important to know how well the existing LDR video compression standards perform for HDR video compression and also identify the challenges for modifying the existing compression standards to account for both LDR and HDR video data. To this end, during the initial phase of HEVC

This work was supported in part by NSERC under Grant STPGP 447339-13 and the ICICS/TELUS People & Planet Friendly Home Initiative at UBC.

TABLE I. SPECIFICATIONS OF THE HDR VIDEO SET

| Sequence | Resolution | Frame Rate (fps) | Number of Frames | Spatial Complexity | Motion Level | Bit Depth |
|---|---|---|---|---|---|---|
| Video1: WalkingGirl | 1920×1080 | 30 | 300 | Low | Low | 12 |
| Video2: WalkingonSnow | 1920×1080 | 30 | 510 | Medium | High | 12 |
| Video3: UBC | 1920×1080 | 30 | 300 | Medium | High | 12 |
| Video4: ICICS | 1920×1080 | 30 | 300 | High | Medium | 12 |

standardization, we evaluated its performance on HDR video using PSNR (peak signal to noise ratio) as an objective quality metric [13]. In this paper, we compare the performance of the standardized version of the HEVC standard with that of the H.264/AVC standard on HDR video content both objectively and subjectively. To this end, a set of HDR videos were captured and a publically available HDR video database, called Digital Multimedia Lab HDR (DML-HDR) dataset, was created. Representative videos of this database were compressed using the HEVC and H.264/AVC video standards. The quality of the resulting compressed streams was evaluated using a prototype HDR display through a series of subjective experiments.

The rest of this paper is organized as follows: Section II describes the quality evaluation process (HDR video set, encoder configurations, HDR display, and subjective evaluations). Section III contains the experiment results and discussions while Section IV concludes the paper.

## II. EXPERIMENT SETTINGS

This section provides details about the DML-HDR video database, the configuration of the HEVC and H.264/AVC encoders, the prototype HDR display system used in our experiment, and the specifications of the subjective experiments.

### A. DML-HDR Video Database

Using an HDR camera (RED Scarlet-X [14]) we captured several indoor and outdoor HDR video sequences. The videos include a wide range of motion, brightness level, and texture complexity. The resolution of the videos is High Definition (HD) at 1920×1080 and the frame rate is 30 fps (frames per second). This HDR video database (DML-HDR) is available to the research community on our website [15].

We chose four representative sequences from the DML-HDR dataset for our experiments. Table I illustrates the specifications of the sequences. A snapshot of the videos is provided in Fig. 1. Note that in Fig. 1, the original HDR frames are tone-mapped to a LDR representation using the adaptive logarithmic mapping [16] for demonstration purposes. The captured videos are originally stored in ".hdr" floating point format and in linear RGB space. However, since the only accepted video format supported by the HEVC and H.264/AVC video encoders is ".yuv" raw files, the videos were converted to 12 bit "4:2:0" raw format using the ITU-R Rec. BT.709 color conversion primaries [17]. To this end, the RGB values are first normalized to [0, 1]. Then, these values are converted to YCbCr color space in accordance to ITU-R Rec. BT.709 [17]. Next, using the separable filter values provided by [18], chroma-subsampling is applied to the YCbCr values to linearly quantize the resulted signals to the desired bit depth

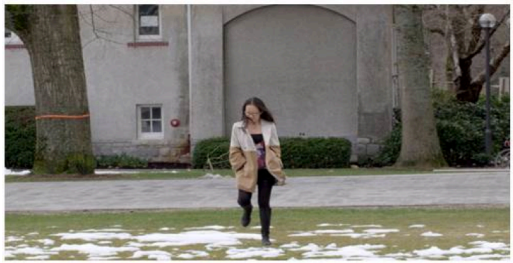
a) Video1, WalkingGirl

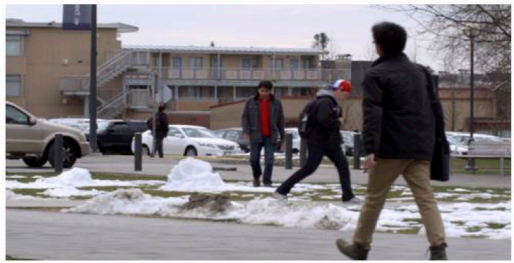
b) Video2, WalkingonSnow

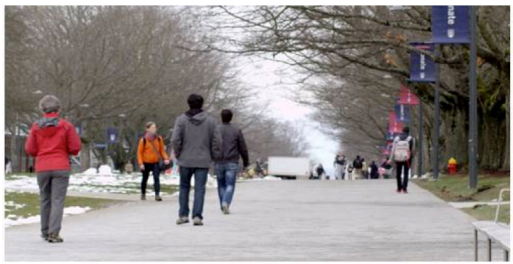
c) Video3, UBC

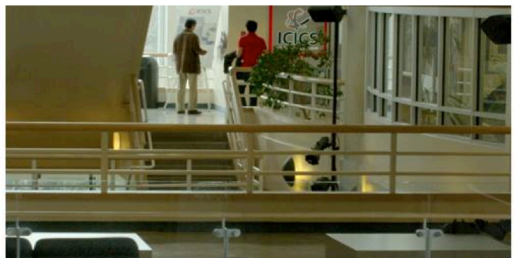
d) Video4, ICICS

Fig. 1. Snapshots of the HDR video sequences

(i.e., 12 bits) [19]. Note that both the HEVC and H.264/AVC video coding standards support videos with up to 14 bit depths.

*B. HEVC and H.264/AVC Encoder Configurations*

To compress the HDR video sequences using the HEVC standard, we used the HM 12.1 reference encoder software [20]. The Random Access High Efficiency configuration with the GOP (Group of Pictures) size of 8 is used for encoding the videos. The SAO, ALF, and RDOQ functions are enabled [11]. The Intra period is set to 32, as the frame rate of the videos is equal to 30 fps [21]. The Quantization Parameter (QP) is set to values of 31, 34, 37, and 40 as it was found empirically that these different values result in a reasonable range of subjective qualities. These QP values have also been used in other studies for HEVC performance evaluations [22].

To compress the sequences using the H.264/AVC video standard, the JM 18.6 encoder software is used [23]. The H.264/AVC encoder is set at High profile with chroma subsampling of 4:2:0, GOP size of 8, frame structure of "IbBbBbBbP", and Intra period of 32 frames. In addition, RDOQ, weighted prediction, and CABAC 8×8 transforms are all enabled. The QP values are set to 27, 30, 33, and 36. These values were found to yield the same bit rates as the QP values chosen for HEVC.

*C. Prototype HDR Display System*

The quality of the compressed HDR videos is evaluated subjectively on a prototype HDR display system. The prototype display is built based on the concept explained in [3]. As illustrated in Fig. 2, this system consists of two main parts: 1) a 40 inch full HD LCD panel in the front, and 2) a projector with HD resolution at the back to provide the backside luminance. The contrast range of the projector is 20000:1. The original HDR video signal is split into two streams, which are sent to the projector and the LCD (see [3] for details). The input signal to the projector includes only the luminance information of the HDR content and the input signal to the LCD includes both luma and chroma information of the HDR video. Using this configuration, the light output of each pixel is effectively the result of two modulations with the two individual dynamic ranges multiplied, yielding an HDR signal. This HDR display system is capable of emitting light at a maximum brightness level of 2700 cd/m$^2$.

*D. Subjective Test Setup*

Subjective experiments are performed to evaluate the quality of the compressed HDR videos. The test material consists of sixteen HEVC-coded and sixteen H.264/AVC-coded sequences resulted from compressing four original HDR videos at four different QP values.

Test session settings are according to ITU-R BT.500 [24]. In each session, a maximum of three subjects participated. Presentation of the test materials is based on the Double Stimulus method, where the original version of the test video is followed by a 3 second gray interval, and the compressed video streams are presented to the viewer. The gray interval is used in order to allow viewers to rest their eyes. After showing the original and compressed videos, a 4-second gray interval is presented, giving time to the viewers to assess the quality of the compressed video against that of the original one and then insert the corresponding mark in the given vertical scale. As the purpose of the subjective test is to evaluate the overall perceived quality (and not the impairment), grading is performed based on the Numerical Categorical Judgment Method [24]. The grading scale consists of 11 distinct quality levels from 0 (worst quality) to 10 (highest quality) [24].

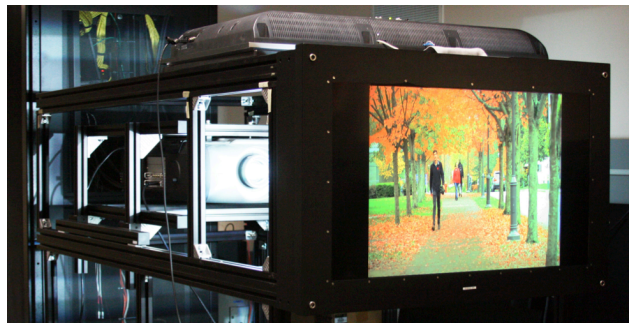

Fig. 2. Illustration of the prototype HDR display system

A total of 17 subjects participated in our experiments. The mean opinion score (MOS) results are calculated after collecting the subjective scores and removing the outliers. Outlier detection is performed in accordance to ITU-R BT.500 [24]. In our tests, one outlier was detected and the subjective data form that participant was discarded.

### III. RESULTS AND DISCUSSIONS

In our study, the performance of the HEVC video coding standard compressing HDR content is compared with that of H.264/AVC, both subjectively and objectively. Fig. 3 shows rate-distortion (RD) curves obtained for each HDR test sequence based on the collected MOS values. As it is observed, HEVC outperforms H.264/AVC by 5.33% to 14.99% (on average 10.18%) in terms of quality, or 14.16% to 38.86% (on average 25.08% savings) in terms of bitrate savings. Fig. 4 demonstrates PSNR-based RD curves for the HDR videos. Note that PSNR is calculated as a weighted average of PSNR values for the luma and chroma components the same way that is described in [22]. For a constant bit rate, HEVC outperforms H.264/AVC by achieving 0.99 dB to 2.45 dB higher PSNR (on average 1.52 dB). At a fixed PSNR value, HEVC provides 21.51% to 41.36% bit rate reduction compared to H.264/AVC (on average 28.46% savings). It is observed that the average distance between RD curves in Fig.3 is less than that in Fig.4. This indicates that the difference between the performance of the HEVC standard and that of H.264/AVC is exaggerated in the case of PSNR-based objective evaluations. This suggests that PSNR is not a representative distortion measure, and instead the compression performance of different codecs should be measured subjectively/perceptually.

In Table II we also list the average bit rate savings and quality improvements achieved by HEVC over H.264/AVC for each HDR test video. It can be observed that, in the case of HDR video content, HEVC outperforms H.264/AVC both subjectively (MOS) and objectively (PSNR).

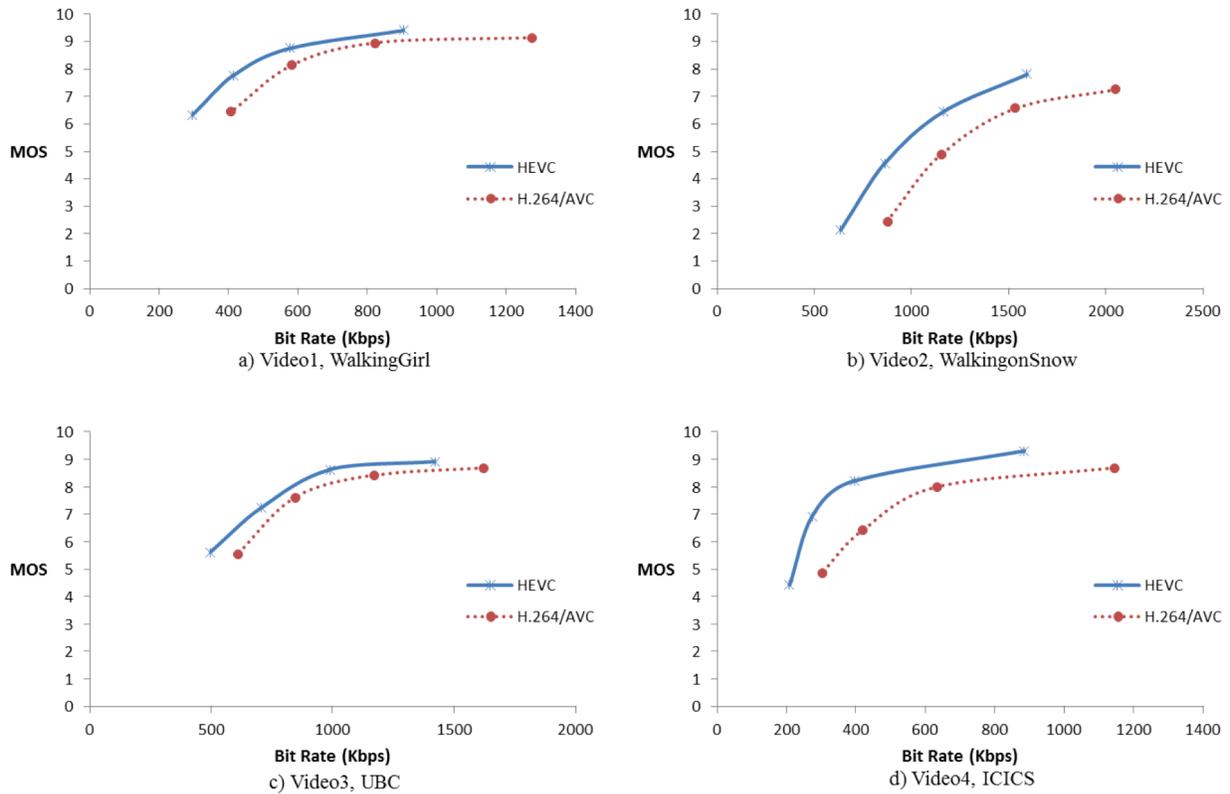

**Fig. 3.** Rate distortion curves for HDR video content: Mean opinion quality score versus bit rate for HEVC and H.264/AVC compression

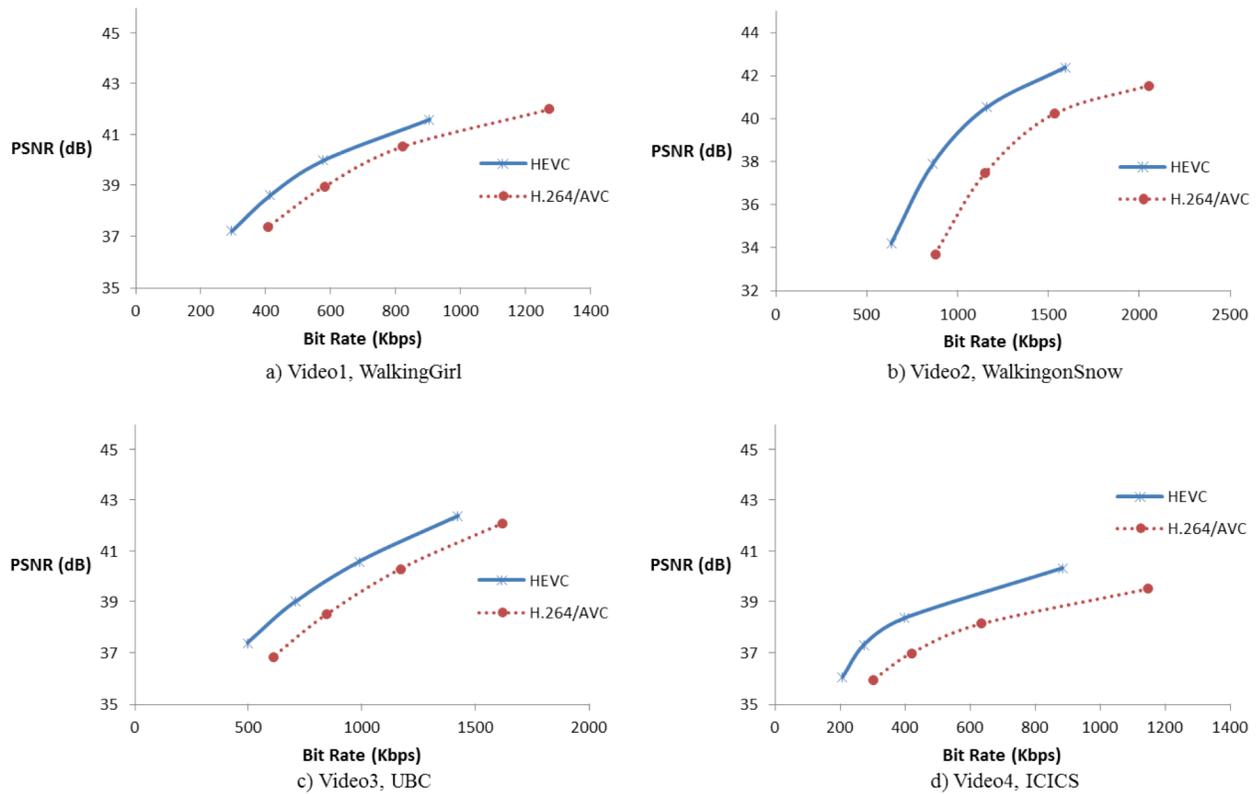

**Fig. 4.** Rate distortion curves for HDR video content: PSNR versus bit rate for HEVC and H.264/AVC compression

TABLE II. AVERAGE IMPROVEMENT ON THE COMPRESSION PERFORMANCE OF HEVC COMPARED TO H.264/AVC

| Sequence | Subjective Evaluations | | Objective Evaluations | |
|---|---|---|---|---|
| | *Average MOS Improvement* | *Average Bit Rate Saving* | *Average PSNR Improvement* | *Average Bit Rate Saving* |
| Video1: WalkingGirl | 6.70% | 24.08% | 0.9908 dB | 21.51% |
| Video2: WalkingonSnow | 14.99% | 23.21% | 2.4534 dB | 28.90% |
| Video3: UBC | 5.33% | 14.16% | 1.2544 dB | 21.97% |
| Video4: ICICS | 13.71% | 38.86% | 1.3942 dB | 41.36% |
| Overall | 10.18% | 25.08% | 1.52 dB | 28.46% |

In summary, our performance evaluation of HEVC and H.264/AVC on HDR content suggests that HEVC has the potential to become the compression standard of choice for both LDR and HDR video applications

IV. CONCLUSION

In this paper we evaluated the performance of the HEVC video coding standard on HDR video content and compared it to that of the H.264/AVC video compression standard on the same content. To this end, we first generated a comprehensive HDR video dataset, containing unique and representative high dynamic range videos (DML-HDR). Then, these videos were encoded using the reference HEVC and H.264/AVC software codecs. The quality of the compressed HDR videos was subjectively and objectively evaluated. Subjective evaluations showed that HEVC provides an average of 25.08% bit rate savings compared to H.264/AVC. Objective evaluations followed a similar trend, showing that HEVC provides an average of 28.46% bit rate savings compared to H.264/AVC.

Our study confirms that the superiority of HEVC standard performance over H.264/AVC standard is not exclusive to the LDR video content and extends to HDR video applications.